\documentstyle[12pt,epsfig]{article}
\begin{document}
\flushbottom

\title{Peripheral Collisions of Relativistic Heavy Ions
}
\author{C. A. Bertulani \\
Physics Department,
Brookhaven National Laboratory,\\
Upton, New York 11973
\footnote{Contribution to
the Symposium on Fundamental Issues in Elementary Matter,
In Honor and Memory of Michael Danos, Bad Honnef, Germany,
September 2000.}}
 \date{\today}
\maketitle

\begin{abstract}
The physics of peripheral collisions with relativistic 
heavy ions (PCRHI) is reviewed. One- and two-photon processes are discussed.
\end{abstract}


\section{Introduction}

Peripheral collisions of relativistic heavy ions 
have attracted a great number of theoretical and
experimental work (see, e.g., \cite{BB94}, and references therein). It is
the purpose of this article to review these phenomena.

\section{Peripheral Collisions}

This field was born in 1924, when E. Fermi had an ingenious idea of relating
the atomic processes induced by fast charged particles (the
electron) to processes induced by electromagnetic waves. In
1934-1935, Weizs\"{a}cker and Williams corrected Fermi's calculation by
including the appropriate relativistic corrections. The original
Fermi's idea is now known as the Weizs\"{a}cker-Williams method \cite{Fe24}, an
approximation widely used in coherent processes in QED and QCD. In this method
the field generated by a fast particle is replaced by a
flux of photons (QED), or a flux of by mesons and gluons (QCD) \cite{BB85}. The number of equivalent photons, $%
n(\omega )$, of given energy, $\omega $, in peripheral collisions of
relativistic heavy ions (PCRHI) can be calculated classically, or
quantum-mechanically. For the electric dipole (E1) multipolarity both
results are identical under the assumption of very forward scattering \cite
{BB85}. In ref. \cite{BB85} the
number of equivalent photons for all multipolarities was calculated exactly. 
It was shown that for the electric dipole multipolarity, E1, the equivalent
photon number, $n_{E1}(\omega)$, coincides with the one deduced by 
Weizs\"acker and Williams. It was also shown that
in the extreme relativistic collisions the equivalent photon numbers for all
multipolarities agree, i.e, $n_{E1}(\omega )\sim n_{E2}(\omega )\sim
n_{M1}(\omega )\sim ...$. The cross sections for one- and two-photon
processes depicted in figure 1(a,b) are given by

\begin{equation}
\sigma _{X}=\int d\omega \frac{n\left( \omega \right) }{\omega }\sigma
_{X}^{\gamma }\left( \omega \right) \;,\;{\rm and}\;\;\sigma _{X}=\int
d\omega _{1}d\omega _{2}\frac{n\left( \omega _{1}\right) }{\omega _{1}}\frac{%
n\left( \omega _{2}\right) }{\omega _{2}}\sigma _{X}^{\gamma \gamma }\left(
\omega _{1},\omega _{2}\right) \;,  \label{epa}
\end{equation}
where $\sigma _{X}^{\gamma }\left( \omega \right) $ is the photon-induced
cross section for the energy $\omega $, and $\sigma _{X}^{\gamma \gamma
}\left( \omega _{1},\omega _{2}\right) $ is the two-photon cross
section. Note that we do not refer to the photon momenta. The virtual
photons are real: $q^{2}=0$, a relation always valid for PCRHI.

Most applications of PCRHI were reviewed in ref. \cite{BB88}.
Since then a great amount of work has been performed in this field. I will
only be able to quote a short number of references.

\subsection{Bremsstrahlung and Delbr\"{u}ck scattering}

Bremsstrahlung (fig. 1c) is a minor effect in PCRHI \cite{BB88}. The cross
section is proportional to the inverse of the square mass of the ions. Most
photons have very low energies (infrared). For 10 MeV photons
the central collisions (CCRHI) deliver 10$^{6}$ more photons than the PCRHI \cite{BB89}.
However, Bremsstrahlung could be relevant to obtain information on the
elastic scattering of photons off unstable particles, like pions: $Z+\pi
\longrightarrow Z+\pi +\gamma $. For a collider the Bremsstrahlung 
cross section is given
by 
\begin{equation}
{d\sigma _{\gamma }\over d\omega} ={16Z^{6}\alpha ^{3}
\over  3\omega
A^{2}m_{N}^{2}}\ln \left( 
{\gamma \over \omega R}\right) 
\end{equation}
 where $%
m_{N} $ is the nucleon mass, $\gamma =2\gamma _{c}^{2}-1$, where $\gamma
_{c} $ is the collider Lorentz gamma factor ($\gamma _{c}\sim 100$ for
RHIC/BNL), and $R$ is the nuclear dimension ($R\sim 2.4\;A^{1/3}$ fm) \cite
{BB89}.
\begin{figure}[thbp]
\nopagebreak
{\epsfxsize=12cm \epsfbox{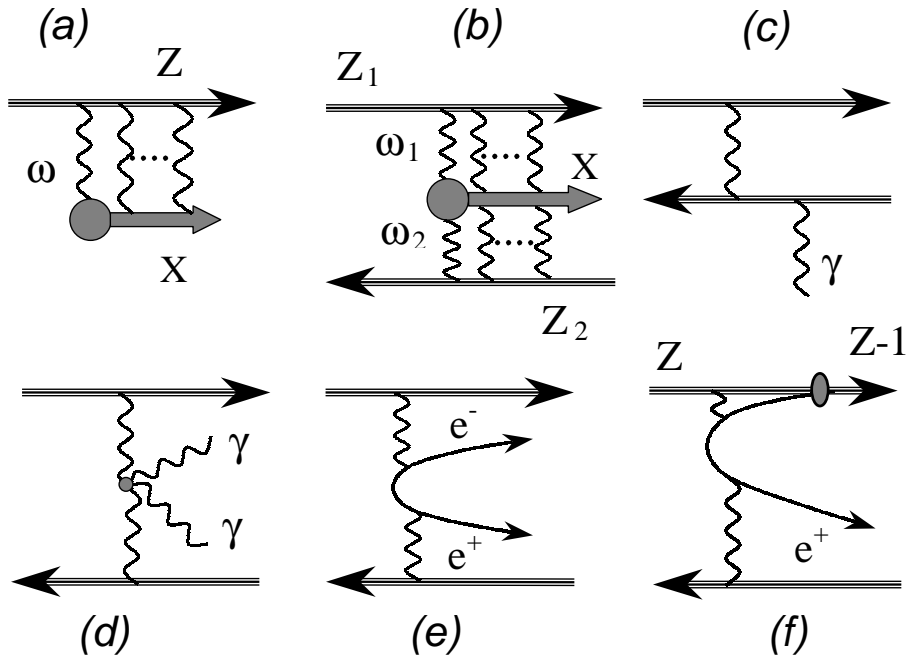}}
\vspace*{10mm}
\caption[]{ PCRHI processes: (a) one-photon, (b) two-photon, (c) Bremsstrahlung,
(d) Delbr\"uck scattering, (e) pair-production, and (f) pair-production with
capture.}
\label{fig1}
\end{figure}

Delbruck scattering ($\gamma ^{\ast }+\gamma ^{\ast }\longrightarrow \gamma
+\gamma $) involves an aditional $\alpha ^{2}$ as compared to pair
production and has never been possible to study experimentally. \ The cross
section is about 50 b for the LHC \cite{BB89} and the process is dominated
by high-energy photons, $E_{\gamma }\gg m_{e}$. A study of this process in
PCRHI is thus promising if the severe background problems arising from CCRHI
can be eliminated. To my knowledge, no experiments of Bremsstrahlung or
Delbr\"{u}ck scattering in PCRHI have been performed so far. The total cross section
for Delbr\"{u}ck scattering ($\omega \gg m_{e}$) in colliders is given by 
\begin{equation}
\sigma _{\gamma \gamma} =2.54Z^{4}\alpha ^{4}r_{e}^{2}
\ln ^{3}\left(
{\gamma \over m_{e}R}\right)   \ \ , 
\end{equation}
where $r_{e}=e^{2}/m_{e}$ is the classical
electron radius \cite{BB89}.

\subsection{Atomic ionization}

Atomic ionization by RHI is used in experiments with
fixed targets for the basic understanding of atomic
structure physics in high-Z few electron atoms such as hydrogen-like or
helium-like uranium atoms. A nice book on this subject has been written by
Eichler and Meyerhof (see also the review by Anholt and Gould) \cite{EM95}.

\begin{figure}[thbp]
\vspace*{10mm}
\begin{minipage}[t]{6.5cm}
{\epsfxsize=6cm \epsfbox{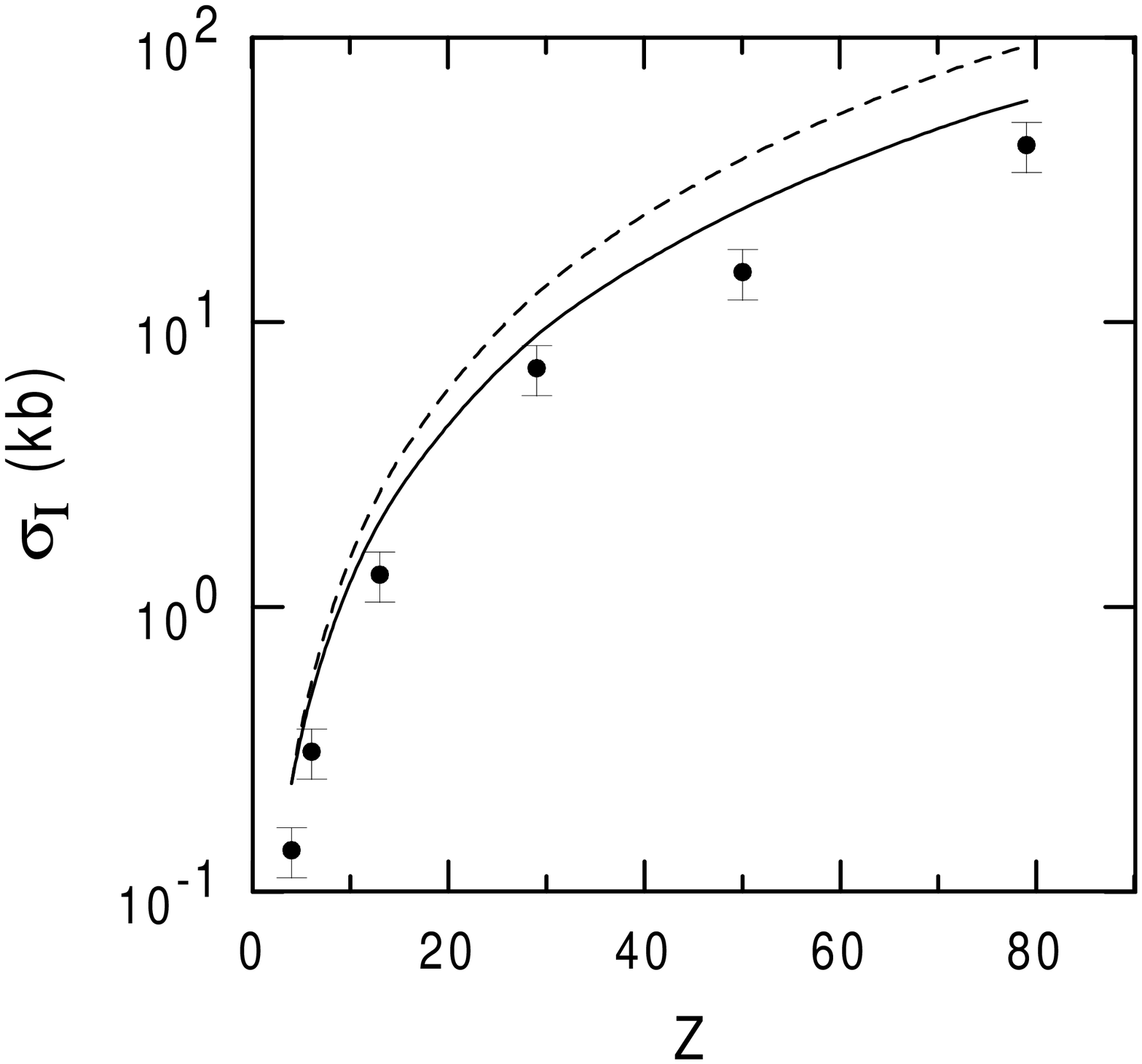}}
\caption[]{\small Atomic ionization cross sections 
for $Pb^{81+}$ (33 TeV) beams on several targets \cite{Kr98}.}
\label{fig2}
\end{minipage} \hfill
\begin{minipage}[t]{6.5cm}
{\epsfxsize=6.1cm \epsfbox{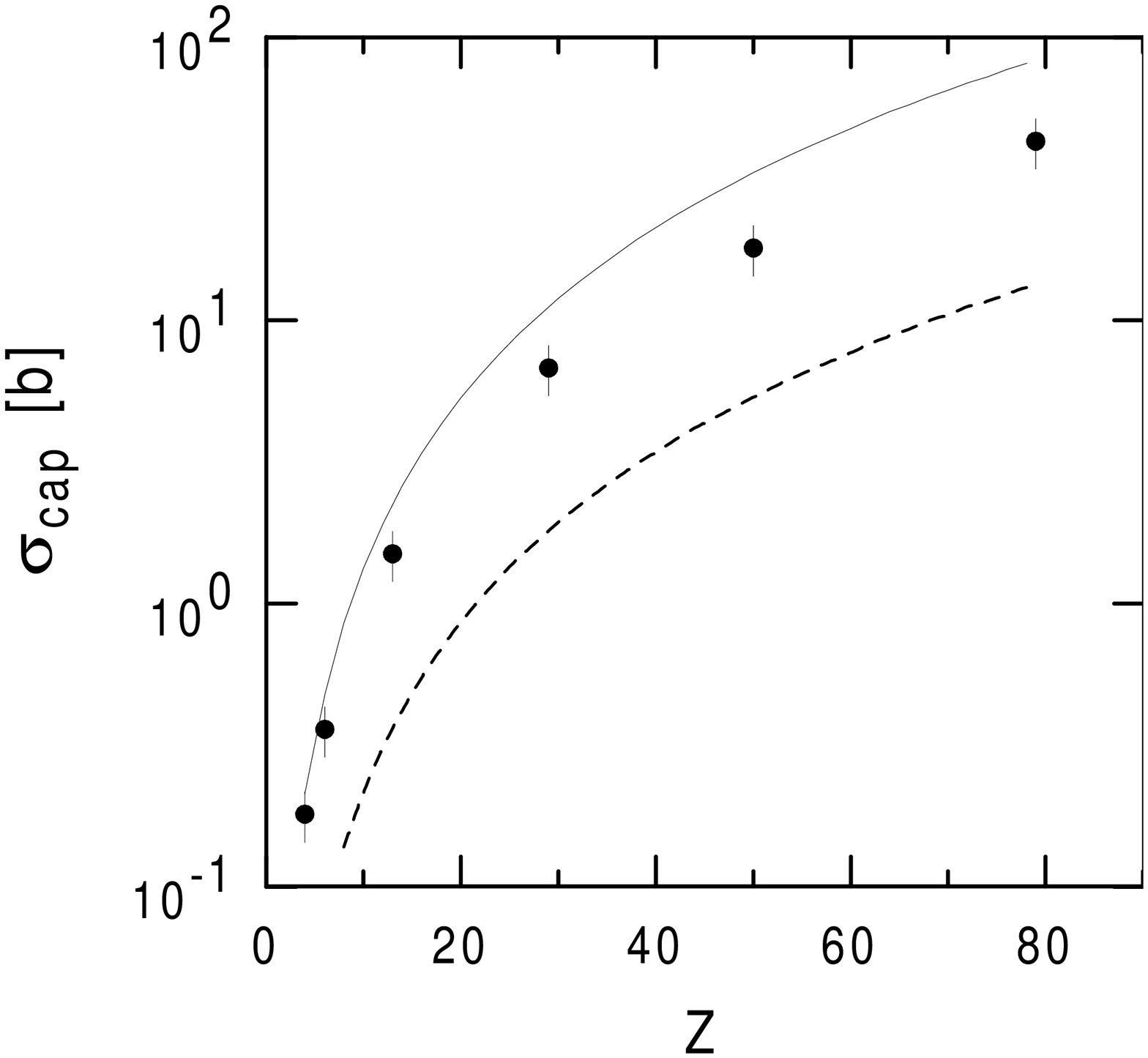}}
\caption[]{\small Pair production with capture 
for $Pb^{82+}$ (33 TeV) beams on several targets \cite{Kr98}.}
\label{fig3}
\end{minipage}
\end{figure}

The cross sections are very large, of order of
kilobarns, increasing slowly with the logarithm of the RHI energy. For a
fixed target experiment using bare naked projectiles one gets \cite{BB88}: 
\begin{equation}
\sigma
_{I}={Z_{P}^{2}r_{e}^{2}\over Z_{T}^2\alpha^2 }\left[ 1.8\pi +9.8\ln
\left( {2\gamma \over Z_{T}\alpha}\right) \right] \ \ , 
\end{equation}
which decreases with the target
charge $Z_{T}$. This is due to the increase of the binding energy of $K$%
-electrons with the atomic charge. \ The first term is
due to close collisions assuming elastic scattering of the electron off the
projectile, while the second part is for distant collisions, with
impact parameter larger than the Bohr
radius. The probability to eject a $K$-electron is much larger than for
other atomic orbitals. \ Recent experiments have reported ionization cross
sections for $Pb^{81+}$ (33 TeV) beams on several targets \cite{Kr98}. In
this case, the role of projectile and target are exchanged in the previous
equation. In figure 2 we show the
results this equation (dashed line) compared to 
the experimental data. Since the targets are
screened by their electrons, the discrepancy is expected. Even the most
detailed calculations by Anholt and Becker \cite{AB87} (solid line) yield
larger cross sections than the experimental data. 

Non-perturbative calculations, solving the time
dependent Dirac equation exactly, were first performed by Giessen and Oak
Ridge groups \cite{Bec83,BS85}. \ It is claimed in ref. \cite{Kr98} that
such non-perturbative calculations do in fact reproduce their data since
they yield \ cross sections which are approximately 70\% the perturbative
ones \cite{Ba91}. However, there are little data 
available for a decisive conclusion about an appropriate theory.

\subsection{Free and bound-free electron-positron pair production}

Between 1933 and 1937, Furry, Carlson, Landau, Lifshitz, Bhabha,
Racah, Nishina, Tomonaga, and several others performed calculations of
$e^+e^-$ production in relativistic collisions of fast
particles (cosmic rays) \cite{Fu33}. The purpose was to test
the newly born Dirac theory for the positron. Starting with the Dirac
equation for the electron and its antiparticle they obtained that,
to leading order in $\gamma $,
\begin{equation}
\sigma ={28\over 27\pi}Z_{P}^{2}Z_{T}^{2}r_{e}^{2}\ln ^{3}\left( 
{\gamma \over
4}\right) \ .
\end{equation}
Unfortunately, in view of the experimental difficulties,  
these results could never been
fully tested. A renewed interest in this process appeared with the
construction of relativistic heavy ion accelerators. For heavy ions with
very large charge (e.g, lead, or uranium) the pair production probabilities and
cross sections are very large. They cannot be treated to first order in
perturbation theory \cite{BB88}, and are difficult to calculate. This resulted
in a great amount of theoretical studies \cite{RB91}. The
formulation of the problem with use of numerical algorithms has varied
wildly among several groups. Semi-analytical approaches have also been used.
The comparison among all these results is rather deceiving, since very
different results are obtained for the production cross sections,
sometimes differing by orders of
magnitude. The perturbative calculations are simple to write down, but
involve rather complicated integrals, specially for low energy electrons,
due to the distortion and relativistic effects on the continuum electronic
wavefunction \cite{EM95}. Screening is also a source of problems. The
non-perturbative calculations are simpler to formulate, but are useless
without a numerical algorithm which contains implicit approximations.

Recently, good developments in more tractable
formulations of the problem appeared in the literature. One replaces the 
Lorentz compressed electromagnetic
fields by delta functions, works with light cone variables, and obtains
almost closed form expressions \cite{Ba97}. However, some of these works have
been strongly criticized \cite{Iv99} because they do not account properly
for Coulomb distortion of the lepton wavefunctions. This
problem was addressed again in ref. \cite{Ei00} with the conclusion that a
full account of distortion of the leptonic wavefunctions has not been
achieved so far. In other words, no theory has ever been possible to tackle
the multiple photon exchange between the electron (or positron) with both the
target and the projectile. A simple thought
reveals why the Coulomb distortion is so important. In
the frame of reference of one of the nuclei, the energy spectrum of the
emitted electron (or positron) is peaked at $\varepsilon \sim 2m_{e}$ \cite
{BB88}. This peak is due to the Coulomb attraction (repulsion)  
which eliminates low energy
components of the leptonic wavefunctions, combined with the decrease in
energy of the number of equivalent photons generated by the other nucleus.
Changing frame of reference, changes this picture. Thus, a correct
calculation should yield two peaks in
the energy distribution of the electron (or positron): one at $\varepsilon
\sim 2m_{e}$, and another around $\varepsilon \sim 2\gamma _{c}^{2}m_e$. 
The
reason it does not appear in perturbative calculations is that the
distortion on the leptonic wavefunctions are calculated with respect to 
only one of the nuclei. Accounting for the distortion  with respect to
both nuclei,
suggests that the total cross section should be about twice the value
obtained by Landau and others \cite{Fu33}. This seems to be a challenge for 
present theoretical calculations. 

An important phenomenon occurs when
the electron is captured in an atomic orbit of the projectile, or of the target.
In a collider this leads to beam losses each time a charge modified
nucleus passes by a magnet downstream \cite{BB89}. A striking
application of this process was the recent production of antihydrogen atoms
using relativistic antiproton beams \cite{Ba96}. Here the positron
is produced and captured in an orbit of the antiproton. Early calculations
for this process
used perturbation theory \cite{AB87,Bec87,BB88}. Evidently, the best way to perform the
calculation is using the frame of reference of the nucleus where the
electron is captured. Many other calculations have been
performed \cite{RB91}. Some of them used non-perturbative approaches, e.g.,
coupled-channels calculations. Initially some discrepancy with perturbative
calculations were found, but later it was shown that non-perturbative
calculations agree with the perturbative ones at the  1\% level (see, e.g., first
reference of \cite{Ba97}). In fact, it would be a surprise if a different
result was found.  The first term of the
perturbation series is already small enough to neglect the inclusion of
higher order terms \cite{BB88}.

The expression 
\begin{equation}
\sigma ={3.3\pi Z^{8}\alpha ^{6}r_{e}^{2}
\over
\exp \left( 2\pi Z\alpha \right) -1}\left[ \ln \left(
0.681\gamma _{c}^{2}\right) -{5\over 3}
\right] 
\end{equation} 
for pair production with electron
capture in PCRHI was obtained in ref. \cite{BB88}. The term $\left[ {...}\right]
^{-1}$ is the main effect of the distortion of the positron wavefunction.
It arises through the normalization of the continuum wavefunctions which
accounts for the reduction of the magnitude of the positron wavefunction
near the nucleus where the electron is localized (bound). Thus, the greater
the $Z$, the less these wavefunctions overlap. The above equation also
predicts a dependence of the cross section in the form 
\begin{equation}
\sigma =A\ln \gamma_{c}+B \ ,
\end{equation}
where $A$ and $B$ are coefficients depending on the system. 
This dependence was used in the analysis of the
experiment in ref. \cite{Kr98}. In recent calculations, attention was given
to the correct treatment of the distortion effects in the positron
wavefunction \cite{He00}. In figure 3 we show the recent experimental
data of ref. \cite{Kr98} compared to the above equation and recent
calculations (second reference of \cite{He00}). These calculations also
predict a $\ln \gamma _{c}$ dependence but give larger cross sections
than in ref.\ \cite{BB88}. The comparison with the experimental data is not
fair since atomic screening was not taken into account. When screening is
present the cross sections will always be smaller by at least a factor 2-4 
\cite{BB88}. The conclusion here is that pair production with electron
capture is a process which is well treated in first order perturbation
theory. The main concern is the correct treatment of distortion effects
(multiphoton scattering) \cite{He00}.

\subsection{Relativistic Coulomb excitation and fragmentation}

Relativistic Coulomb excitation is becoming a popular tool for the
investigation of the intrinsic nuclear dynamics and structure of the
colliding nuclei \cite{Au98}. This is specially important in reactions
involving radioactive nuclear beams \cite{BCH93}. 
Coulomb excitation and dissociation of such nuclei are 
common experiments in this field \cite{BB94,AB95}. The advantage is
that the Coulomb interaction is very well known. The real
situation is  more complicated since the contribution of the
nuclear-induced processes cannot be entirely separated in the experimental
data.  
The treatment of the dissociation problem by
nuclear forces is very model dependent, based
on eikonal or multiple Glauber scattering approaches \cite{BCH93,BBK92}. Among
the uncertainties are the in-medium nucleon-nucleon cross sections at
high-energies, the truncation of the multiple scattering process and the
separation of stripping from elastic dissociation of the nuclei \cite{HRB91}.
Nonetheless, specially for the very weakly-bound nuclei, relativistic
Coulomb excitation has lead to very exciting new results \cite{BCH93,BBK92}.

\begin{figure}[thbp]
\vspace*{10mm}
\begin{minipage}[t]{6.5cm}
{\epsfxsize=6.5cm \epsfbox{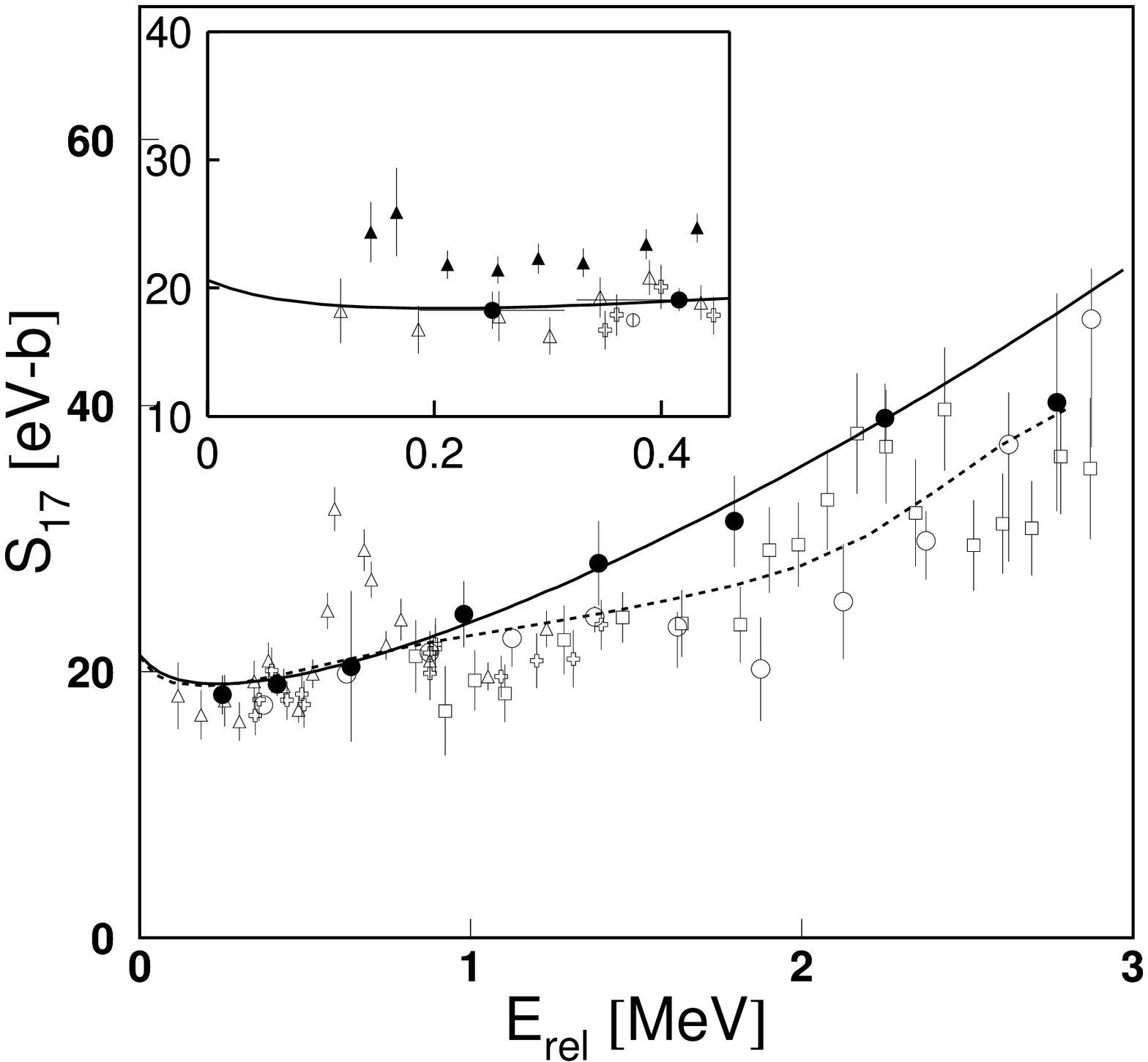}}
\caption[]{\small S-factors for the $^7Be(p,\gamma)^8B$ reaction.}
\label{fig4}
\end{minipage} \hfill
\begin{minipage}[t]{6.5cm}
{\epsfxsize=6.5cm \epsfbox{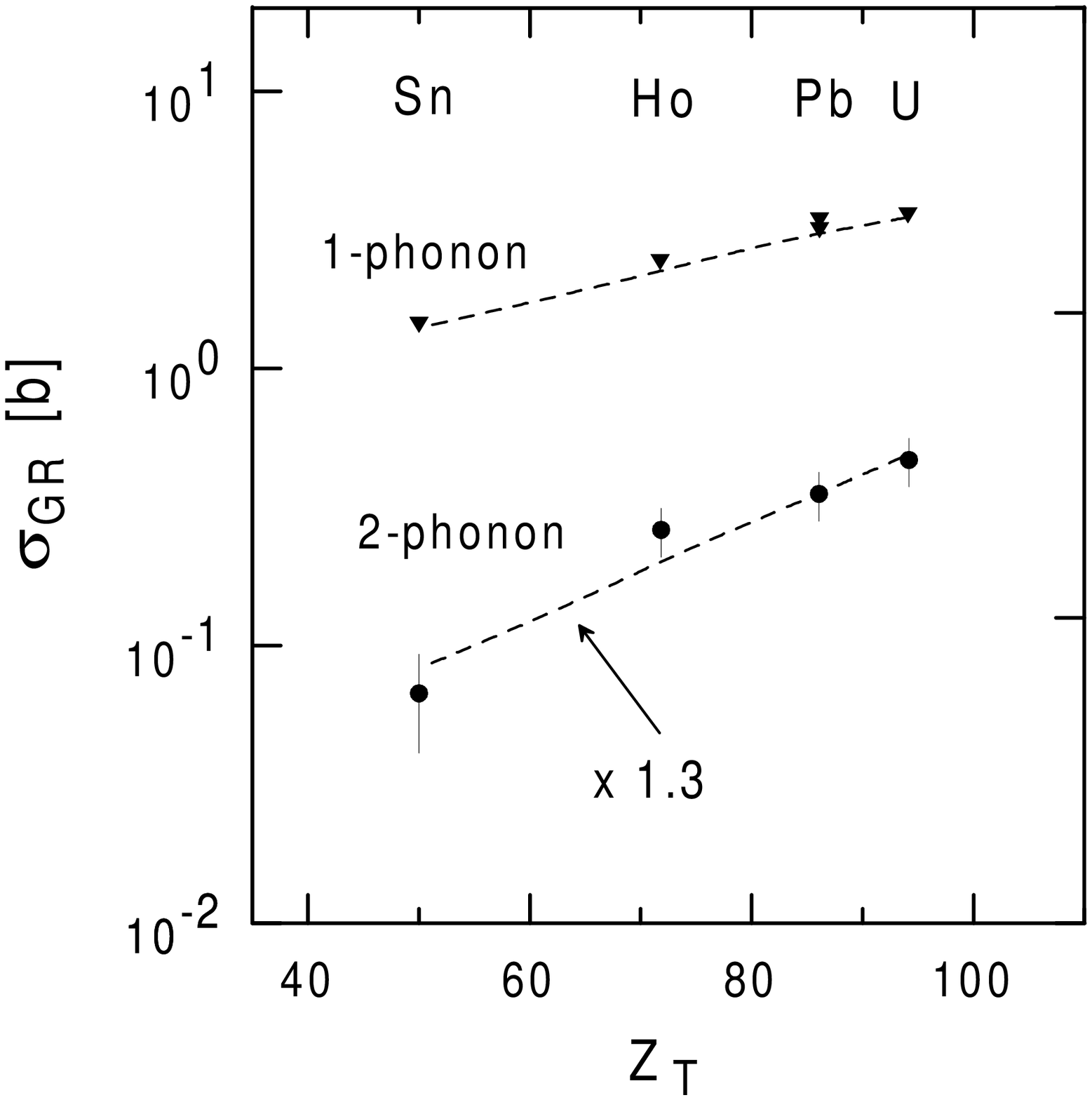}}
\caption[]{\small Cross sections for the excitation of the GDR and
the DGDR.}
\label{fig5}
\end{minipage}
\end{figure}

The Coulomb breakup of $^{11}Li$\ has lead to interesting results which gave
rise to a series of speculations about the reaction mechanism, the
dynamics, and the structure of this nucleus. One speculates if
the reaction proceeds under a single or multiple photon-exchange between the
projectile and the target. In the first case, perturbation theory gives a
straight relation between the data and the matrix element for
electromagnetic dissociation. Such matrix elements are the clearest probes
one can get about the nuclear structure of these nuclei. In the second case,
often called by post-acceleration effects \cite{BBK92}, one has to perform a
non-perturbative treatment of the reaction what complicates the extraction
of the electromagnetic (mainly E1) matrix elements. 
By solving such
problems one expects to learn if
the Coulomb-induced breakup proceeds via a resonance or by the direct
dissociation into continuum states  \cite{BBK92}. There is a strong ongoing effort
to use the relativistic Coulomb excitation technique also for studying the
excitation of bound excited states in exotic nuclei, to obtain 
information on gamma-decay widths, angular momentum,
parity, and other properties of hitherto unknown states \cite{AB95}.

Another application of Coulomb dissociation of radioactive
nuclei is in astrophysics. Radiative capture reactions are known to play a
major role in astrophysical sites, e.g., in a pre-supernova \cite{Rolfs}.
Some of these reactions, like for example, $^{7}Be\left( p,\gamma \right)
^{8}B$, can be studied via the inverse photo-dissociation reaction $%
^{8}B\left( \gamma ,p\right) ^{7}Be$. 
One often uses the astrophysical S-factor, defined by 
\begin{equation}
S(E)=E\sigma \left( E\right) \exp \left[ -2\pi \eta \left( E\right) \right] 
\ , \ \ \ \ 
{\rm where} \ \ \  \eta \left( E\right) ={Z_{1}Z_{2}e^{2}\over 
\hbar \sqrt{2\mu _{12}E}} \ ,
\end{equation}
where $E$ is the relative kinetic energy of the two nuclei.
The matrix
elements involved in the dissociation processes are the same as those
involved in the absorption by real photons \cite{BB88}. 
One of the
experiments using this technique was performed at the GSI/Darmstadt \cite
{Iw99}. The S-factor obtained in this experiment is shown in figure 4 as
solid circles. Such experiments are very important specially in those cases
where the radiative capture cross section is so small that the direct fusion
experiments are very difficult, or even impossible to carry out. The
contribution of the nuclear-induced breakup and of post-acceleration effects
also limit the use of the Coulomb dissociation technique for this purpose.
But, in view of the much more difficult experiments on direct capture, it is
a very useful alternative method.

Another important application of relativistic Coulomb excitation is the
study of the DGDR (Double Giant Dipole Resonance). A GDR occurs in nuclei at
energies of 10-20 MeV. Assuming that these are harmonic vibrations of 
protons against neutrons, one expects that DGDRs, i.e., two giant dipole
vibrations superimposed in one nucleus, will have exactly
twice the energy of the GDR \cite{BB88,Au98}. Small deviations are expected
from non-harmonic properties of the nuclear response. A series of
experiments at the GSI/Darmstadt have obtained energy spectra, cross
sections, and angular distribution of fragments following the decay of the
DGDR. 
Initially they observed cross sections twice as large as
expected from theoretical calculations. These results lead to a series
of studies on deviations from the harmonic picture of the giant
resonances. More recently, new experiments and new analysis have shown that
the experimental cross sections are only about 30\%, or less, bigger than
the theoretical ones. This is shown in figure 5 where the cross sections
for the excitation of 1-phonon (GDR), 
\begin{equation}
\sigma _{1}\sim 2\pi S\ln \left[
{2\gamma A_{T}^{1/3}\over  A_{P}^{1/3}+A_{T}^{1/3}}\right]  \ ,
\end{equation}
while for the 2-phonon state it is 
\begin{equation}
\sigma \sim {S^{2}\over\left(
A_{P}^{1/3}+A_{T}^{1/3}\right) ^{2}}\ , \ \ \ \ 
{\rm where} \ \ \ S= 5.45\times
10^{-4}{Z_{P}^{2}Z_{T}N_{T}\over A_{T}^{2/3}}\ \  {\rm mb}
\ .
\end{equation} 
The dashed lines of figure 5
are the result of more elaborate calculations \cite{Au98}. The GSI
experiments are very promising for the studies of the nuclear response in
very collective states. One should notice that after many years of study
of the GDRs and other collective modes, the width of these states are still
poorly explained theoretically, even with the best microscopic approaches
known sofar. The extension of these approaches to the study of the width of
the DGDRs will be helpful to improve such models. Hopefully, after
addressing more fundamental questions on QCD and QED of strong
electromagnetic fields, some experiments on relativistic colliders will also
be proposed for the study of nuclear structure issues, specially the issue
of the DGDR.

The DGDR contributes only to about 10\% of the total fragmentation cross
section induced by Coulomb excitation with relativistic heavy ions. The main
contribution arises from the excitation of a single GDR, which decays mostly
by neutron emission. This is also a potential source of beam loss in relativistic
heavy ion colliders \cite{BB94}, and an important fragmentation mode of 
relativistic nuclei in cosmic rays.

\subsection{Meson and hadron production}

The production of heavy lepton pairs ($\mu^+\mu^-$, or $\tau^+\tau^-$), or of
meson pairs (e.g., $\pi^+\pi^-$) can be calculated using the second of equation 
(1). One just needs the cross sections for $\gamma\gamma$ production of these pairs.
Since they depend on the
inverse of the square of the particle mass \cite{BB89},
the pair-production cross sections are much smaller in this case. 
The same applies to single meson production by $\gamma\gamma$ fusion. The
$\gamma\gamma$ cross section is given by 
\begin{equation}
\sigma_{\gamma\gamma\rightarrow M}= 8 \pi^2
(2J+1){\Gamma_{M\rightarrow\gamma\gamma}\over M}\delta(W^2-M^2) 
\ ,
\end{equation}
where $J$, $M$, and 
$\Gamma_{M\rightarrow\gamma\gamma}$ are the spin, mass and two-photon decay width of the
meson, $W$ is the c.m. energy of the colliding photons \cite{BB89}. A correction for
the equivalent photon numbers is necessary, since the two-photon energy folding in eq. (1) 
has to account for the space geometry of the two-photon collision \cite{BF90}. 

A careful study of the production of meson pairs and single mesons in PCRHI was
performed recently in ref. \cite{RN00}. 
In table I we show the magnitude of the cross sections for single meson production 
at RHIC and at LHC \cite{RN00}. Also shown are the cross sections due to difractive
processes (pomeron-pomeron exchange). We see that they are several orders of
magnitude smaller than those from $\gamma\gamma$ fusion. The cross sections for the
production of $\eta_c$, $\eta_c'$ and $\eta_b$ are very small due to their higher masses.
Similar studies have been done for meson production in $\gamma$-nucleus interactions.
Particles like $\Delta$, $\rho$, $\omega$, $\phi$, $J/\Psi$, etc, can be produced in this
way \cite{BHT98}.  

The possibility to produce a Higgs boson via $\gamma\gamma$ fusion was suggested in
ref. \cite{Gr89}. The cross sections for LHC are of order of  1 nanobarn, about the 
same as for
gluon-gluon fusion. But, the two-photon processes can also produce $b\bar{b}$ pairs which
create a large background for detecting the Higgs boson. A good review of these topics
was presented in ref. \cite{BHT98}.

\begin{table}[thb]
\vspace*{-10pt} 
\caption{ Particle production in PCRHI at RHIC and at LHC. Masses are in MeV,
decay widths in keV, and cross sections in mb. The cross sections are for
$\gamma\gamma$ and pomeron-pomeron (${\cal PP}$) exchange processes, respectively.}
\vspace*{-12pt} 
\par
\begin{center}
\begin{tabular}{lllllll}
\hline
&  &  &  &  \\[-10pt] 
Meson & M & $\Gamma_{X\rightarrow\gamma\gamma}$ & $RHIC_{\gamma\gamma}$ & $%
LHC_{\gamma\gamma}$ & $RHIC_{\cal PP}$ & $LHC_{\cal PP}$\\ \hline
&  &  &  &  \\[-10pt] 
$\pi^0$ & 135 & $8 \times 10^{-3}$ & 7.1 & 40 & 0.05 & 0.367 \\ 
$\eta$ &  547 & 0.463 & 1.5 & 17 & 0.038 & 0.355 \\ 
$\eta'$ & 958 & 4.3 & 1.1 & 22 & 0.04 & 0.405 \\  \hline
\end{tabular}
\end{center}
\end{table}

\section*{Acknowledgment(s)}

The author is currently a fellow of the John Simon Guggenheim Memorial Foundation.
This work was partially supported by the Brazilian funding agencies CNPq,
CAPES, FUJB and MCT/FINEP/CNPQ(PRONEX) under contract No. 41.96.0886.

\end{document}